\newcount\subeqnno
\def\bse{\begingroup
\refstepcounter{equation}
\subeqnno=\arabic{equation}
\setcounter{equation}{0}
\def\theequation{\the\subeqnno\alph{equation}}
}
\def\ese{\setcounter{equation}{\the\subeqnno}\endgroup}

\newcommand{\mb}[1]{\ifmmode#1\else\mbox{$#1$}\fi}

\newcommand\al{\mb{\alpha}}

\newcommand\be{\mb{\beta}}
\newcommand\ga{\mb{\gamma}}

\newcommand\ep{\mb{\epsilon}}

\newcommand\et{\mb{\eta}}

\newcommand\la{\mb{\lambda}}
\newcommand\Ph{\mb{\Phi}}
\newcommand\Ps{\mb{\Psi}}

\newcommand\calC{\mb{{\cal C}}}

\newcommand\calG{\mb{{\cal G}}}
\newcommand\calH{\mb{{\cal H}}}

\newcommand\calL{\mb{{\cal L}}}
\newcommand\calM{\mb{{\cal M}}}

\newcommand\calV{\mb{{\cal V}}}

\newcommand{\cd}{\partial}

\newcommand{\x}{\mb{\times}}
\newcommand{\beq}{\begin{equation}}
\newcommand{\eeq}{\end{equation}}
\newcommand{\nn}{\nonumber}
\newcommand{\bea}{\begin{eqnarray}}
\newcommand{\eea}{\end{eqnarray}}
                                       
\newcommand{\fn}{\footnote}
\newcommand{\norm}[1]{\parallel \! {#1} \! \parallel}

\newcommand{\emb}{\mb{{\rm emb}}}
\newcommand{\Ad}{\mb{{\rm Ad}}}

\newcommand{\tr}{\mb{\rm Tr}}
\newcommand{\deriv}[2]{\frac{d {#1}}{d {#2}}}

\newcommand{\pderiv}[2]{\frac{\cd {#1}}{\cd {#2}}}

\newcommand{\m}{\mb{\calM}}
\newcommand{\inprod}[2]{\langle {#1}, {#2} \rangle}
\newcommand{\inproda}[2]{\{ {#1}, {#2} \}}

\newtheorem{theorem}{Theorem}

\newtheorem{conjecture}[theorem]{Conjecture}
\newcommand{\proof}[1]{\smallskip \noindent {\bf Proof} \ \ {#1} \ \
$\Box$ \medskip}
\newcommand{\result}[2]{\medskip \noindent {\bf {#1}}{\it
\ {#2}} \medskip} 

\documentstyle[12pt]{article}
\topmargin = - 0.5 cm
\begin{document}
%\bibliographystyle{unsrt}

%the text input

%%%%%%%%%%%%       abstract and title page    %%%%%%%%%%%%%%%%%%%%%%

\begin{flushright}
CERN-TH/95-185 \\
DAMTP-95-06
\end{flushright}
\begin{center}
\LARGE{EMBEDDED VORTICES\\}
\vspace*{0.5cm}
\large{Nathan\  F.\  Lepora
\fn{e-mail: N.F.Lepora@amtp.cam.ac.uk}$^{,1}$
and Anne-Christine\ Davis
\fn{e-mail: A.C.Davis@amtp.cam.ac.uk}$^{,1,2}$\\}
\vspace*{0.2cm}
{\small\em 1) Department of Applied Mathematics and Theoretical
Physics,\\
University of Cambridge, Silver Street,\\
Cambridge, CB3 9EW, U.\ K.\\ }
\vspace*{0.2cm}
{\small\em 2) Theory Division, CERN,\\
Geneva 23, CH-1211, Switzerland.\\} 
\vspace*{0.2cm}

{July 1995}
\end{center}

\begin{abstract}
We present a discussion of embedded vortices in general
Yang-Mills theories. The origin of a family structure of solutions is
shown to be group theoretic in nature and a procedure for its
determination is developed. Vortex stability can be quantified into
three types: Abelian topological stability, non-Abelian topological
stability, and dynamical stability; we relate these to the 
family structure of vortices, in particular discussing how Abelian
topological and dynamical stability are related. The formalism
generally encompasses embedded domain walls and embedded monopoles
also.   
\end{abstract}
\thispagestyle{empty}
\vskip 90pt
CERN-TH/95-185\\
July 1995
\newpage
\setcounter{page}{1}

%%%%%%%%%%%%       introduction   %%%%%%%%%%%%%%%%%%%%%%%%%%%%%%%%%

\section{Introduction}
\label{sec-1}

In this paper we discuss how one determines the embedded vortex
structure of a general Yang-Mills theory. The concept of an embedded
defect, in a proper mathematical context, was introduced by Barriola,
Vachaspati and Bucher~\cite{Vach94}; they consider embedded defects of
the general form:
\begin{center}
(radial profile function) $\x$ (angular exponential of Lie subalgebra).
\end{center}
The usual topological defects therefore fit into this class. However,
one should also be aware that there are other defect solutions not of
this form. We aim to discuss the spectrum and properties of vortex
solutions of the embedded defect form above.

One should note that many examples of the formalism contained within
this paper are given in a companion paper, which also contains the
usual historical perspective~\cite{me3}.

\subsection*{Prenote: Restriction to Formalism}

Because of extra complications for a variety of symmetry breaking
schemes we shall make the following restriction on theories considered
in this paper: 
\begin{quote}
For symmetry breaking schemes of the form $G \rightarrow H$,
write $\calG = \calH \oplus \calM$ and split $\calM$ into irreducible
parts under $\Ad(H)$:
$\calM=\calM_1 \oplus ... \oplus \calM_n$.

Then we confine our discussion to those $\calM_i$'s such that
\[
{\rm{rank}}(\calM_i) = 1
\]
where the rank of $\calM_i$ is defined to be the dimension of
the maximal Abelian subalgebra within $\calM_i$.
\end{quote}
The terminology in the above is explained within the paper. However,
we feel it should be stated here so that one is aware of this
restriction from the start.

%%%%%%%%%%%%       Section 2      %%%%%%%%%%%%%%%%%%%%%%%%%%%%%%%%%

\section{Embedded Vortices}
\label{sec-2}

In order to provide an understanding of the nature and stability of
embedded defects it is necessary  to have a framework for describing
them.  This has been provided by Barriola, Vachaspati and Bucher
\cite{Vach94}. For work in this paper it is necessary to use the 
underlying group theory behind the theory of embedded defects. Thus,
in this section we indicate how group theory connects with the
formalism introduced in~\cite{Vach94}.

Consider a Yang-Mills field theory with gauge symmetry group $G$
that breaks to a smaller group $H$ via condensation of a scalar
field $\Ph$ taking values in a vector space $\calV$. One describes the
action of $G$ upon $\calV$ by the representation, $D$ of
$G$. The corresponding Lie algebra, $\calG$, acts upon
$\calV$ by the derived representation $d$, related to $D$ by
$D(e^X)=e^{d(X)}$. 

Such a field theory is specified by a Lagrangian describing the
interaction of a gauge field $A^\mu \in \calG$ with the scalar field
$\Phi$ 
\beq
\calL[\Phi, A^\mu]= -\frac{1}{4} \inprod{F_{\mu \nu}}{F^{\mu \nu}}
+ \frac{1}{2} \inprod{( \partial_\mu  + d(A_\mu)) \Phi}{( \partial^\mu
+ d(A^\mu))\Phi} - V[\Phi],
\eeq
with 
\beq
\label{field tensor}
F_{\mu \nu} = \partial_\mu A_\nu - \partial_\nu A_\mu + [A_\mu,
A_\nu],
\eeq
and $V$ the scalar potential. To give $\calL$ the necessary symmetry
properties the inner products $\inprod{.}{.}$ are constrained to be
invariant under the action of $G$. Note that we are using the same
symbol to denote both inner products; we hope it should be clear from
the context which we are using.

The inner product upon $\calV$ is defined to be of the real form
\beq
\label{ip}
\inprod{\Phi}{\Psi} = {\rm Re}(\Phi^\dagger \Psi),
\eeq
for invariance of the scalar field kinetic term under the group
$Gl(\calV)$.

The inner product upon $\calG$ is defined to 
\beq
\label{scaleip}
\inprod{.}{.} = \frac{1}{q_1^2} \inproda{.}{.}_1 + ... +
\frac{1}{q_n^2} \inproda{.}{.}_n,
\eeq
with $\inproda{.}{.}_i$ the inner product $\inproda{X}{Y}=-p\tr(XY)$, 
restricted to $\calG_i$, one of the mutually commuting subalgebras of 
$\calG=\calG_1 \oplus ... \oplus \calG_n$. The $n$ real scales
$\{q_i\}$ are related to the gauge coupling constants of
the theory. This inner product (\ref{scaleip}) is the most general
$\Ad(G)$-invariant inner product, where $\Ad$ is the adjoint
representation $\Ad(g)X=gXg^{-1}$ for $X \in \calG$.

A minima of $V$, say $\Phi_0 \in \calV$, is arbitrary because of the
degeneracy of choice given by the vacuum manifold $M=D(G)\Ph_0$. The
vacuum manifold $M \cong G/H$, with $H$ the residual symmetry group
determined by $\Ph_0$ to be $H = \{g \in G : D(g) \Ph_0 = \Ph_0
\}$. Then the inner product (\ref{scaleip}) defines the decomposition 
\beq
\calG = \calH \oplus \m,
\eeq
with $\calH$ the Lie algebra of $H$.

To construct an {\em embedded defect solution} on such a field theory
one chooses a subgroup $G_{\emb} \subset G$ such that the
corresponding homotopy group $\pi_k (G_\emb / H_\emb)$ is non-trivial,
where $H_\emb = G_\emb \cap H$. The crucial idea is to find a smaller
theory on which a defect is topological and then extend this solution
back to the full theory. This gives rise to the concept of an {\em
embedded subtheory}: a pair $(\calV_\emb, G_\emb) \subset \calV \x G$,
with $\calV_\emb$ a minimal non-trivial vector subspace of
$\calV$ invariant under $D(G_\emb)$.  

From the inner products (\ref{ip}, \ref{scaleip}) we identify
$\calV_\emb^\perp$ and $\calG_\emb^\perp$ as the corresponding
orthogonal subspaces to $\calV_\emb$ and $\calG_\emb$ within $\calV$
and $\calG$ respectively.

The main result from \cite{Vach94} is that a defect topological in the
embedded subspace 
\bse
\bea
\label{eq-1a}
\Ph(x) &=& \Ph_\emb(x) \in \calV_\emb, \\
\label{eq-1b}
A^\mu(x) &=& A^\mu_\emb(x) \in \calG_\emb,
\eea
\ese
remains a solution of the full theory provided the following
conditions hold: 
\begin{enumerate}
\item
The scalar potential satisfies
\beq
\label{scderiv}
\pderiv{V}{\Ps}[\phi] := \inprod{\Psi}{\pderiv{V}{\Phi}[\phi]}= 0,
\ {\rm for}\  \Ps \in \calV_\emb^\perp, \phi \in \calV_\emb, \Phi \in
\calV. 
\label{eq-2}
\eeq
\item 
The scaled representation satisfies
\beq
d(X^\perp)\Ph \in \calV_\emb^\perp,\ {\rm for}\  \Ph \in \calV_\emb \
{\rm and}\ X^\perp \in \calG_\emb^\perp.
\label{eq-3}
\eeq
\end{enumerate}
These two conditions are provided by a direct substitution
of~(\ref{eq-1a}, \ref{eq-1b}) into the field equations, requiring that 
`embedded fields do not induce non-embedded currents'~\cite{Vach94}.

We now specialise to vortex solutions. In doing so we will shift the
emphasis away from defining the embedded vortex by the pair
$(\calV_\emb, G_\emb)$; instead we consider the equivalent pair
$(\Ph_0 , X ) \in \calV_\emb \x \calG_\emb$. 

An embedded vortex defined by the pair $(\Ph_0, X)$ is of the
form
\bse 
\bea 
\label{vor1}
\Ph(r,\theta) &=& f_{\rm NO}(X; r) \exp(d(X) \theta)
\Ph_0,\\ 
\label{vor2}
A(r,\theta) &=& \frac{g_{\rm NO}(X; r)}{r} X
\underline{\hat{\theta}}. 
\eea 
\ese 
The two profile functions $f_{\rm NO}$ and $g_{\rm NO}$ depend 
upon the generator of the embedded defect, and we discuss how
so in the appendix. One also has the single valuedness constraint that
\beq 
\exp(2 \pi d(X)) \Ph_0 = \Ph_0.  
\eeq

The above vortex~(\ref{vor1}, \ref{vor2}) represents the
minimum energy configuration of this form for $X\in \cal M$,
since if we were to add any component in $\cal H$ the effect would be
to leave the scalar field unchanged but to add an extra term to the
gauge-field magnetic energy. Thus we restrict $(\Phi_0, X) \in
\calV \x \calM$.

It is conceptually important to relate the above description of
embedded vortices to the formalism of embedded defects. There is a
natural correspondence between defining a vortex from the pair
$(\calV_\emb, \calG_\emb)$ to the pair $(\Ph_0, X)$ provided by
the following relations:
\bse
\bea 
\calG_\emb &=& {\bf R} X,\\
G_\emb &=& \exp({\bf R} X),\\
\calV_\emb &=& {\bf R}  D(G_\emb) \Ph_0.
\eea
\ese
with ${\bf R}$ the real number line.

Applying Eqs.~(\ref{eq-2}, \ref{eq-3}) for existence of embedded
defects to the above description of embedded vortices we see that
(\ref{eq-2}) is satisfied provided that a suitable potential is
chosen; this condition is essentially a restriction on the form of the
field theory. The condition (\ref{eq-3}) is more fundamental; we prove
below that~(\ref{eq-3}) is equivalent to the condition that $X$ be a
vortex generator provided
\beq
\langle d(X)\Ph_0, d(X^\perp)\Ph_0 \rangle = 0,
\label{eq-10}
\eeq 
for $\inprod{X^\perp}{X}=0$.
This has the rather nice interpretation that only subspaces over
which the derived representation is {\em conformal} are admissible to
define embedded vortices. {\em I.e.} viewing $d(X)\Ph_0$ as a map
between $\calG$ and $\calV$, Eq.~(\ref{eq-10}) 
says that this map must preserve orthogonality with respect
to~(\ref{ip}) from the domains $\calG_\emb$ and $\calG_\emb^\perp$.

We now prove the equivalence of conditions (\ref{eq-3}) and
(\ref{eq-10}): writing $\calV_\emb = {\bf R}\Phi \oplus {\bf
R}d(X) \Phi$ we see that for $\Phi \in \calV_\emb$ and
$\inprod{X^\perp}{X}=0$,
\bea
d(X^\perp) \Phi \in \calV_\emb^\perp
&\Leftrightarrow&
\langle \calV_\emb , d(X^\perp) \Ph \rangle =0 \nn \\
&\Leftrightarrow&
\langle d(X) \Phi , d(X^\perp) \Ph \rangle =0 \nn \\
&\Leftrightarrow&
\langle d(X)\Ph_0, d(X^\perp)\Ph_0 \rangle = 0. \nn
\eea
The second step by virtue of $\inprod{d(\calG)\Phi}{\Phi}=0$, and the 
last by virtue of for general $\Phi \in \calV_\emb$ there
exists a $g \in G_\emb$ which rotates the above $\Phi$ to a constant
times $\Phi_0$. 
This transformation leaves $\calG_\emb$ and $\calG_\emb^\perp$
invariant, because $G_\emb$ is a group.

%%%%%%%%%%%%       section 3      %%%%%%%%%%%%%%%%%%%%%%%%%%%%%%%%%

\section{Families of Embedded Vortices}
\label{sec-3}

We will now determine the family structure of embedded vortices in a
general Yang-Mills theory. This essentially boils down to a
description of group actions of $G$ upon the vacuum manifold. To
describe this properly requires some rather technical work, and so, for 
pedagogical reasons, we shall assume some of the more technical points
and cover a proper, more formal treatment in another paper~\cite{me4}. 

In summary of the previous section: an embedded vortex is defined by
the pair $(\Ph_0, X)$ to be  
\bse
\bea
\label{eq-11a}
\Ph(r,\theta) &=& f_{\rm NO}(X;r) \exp(d(X) \theta) \Ph_0,\\
\label{eq-11b}
A(r,\theta) &=& \frac{g_{\rm NO}(X;r)}{r} X
\underline{\hat{\theta}}, 
\eea
\ese
which is a solution provided
\bse
\bea
\pderiv{V}{\Ps}{[\Ph]} &=& 0,\ {\rm for}\  \Ps \in \calV_\emb^\perp,
\Ph \in \calV_\emb, \\ 
\inprod{d(X)\Ph_0}{d(X^\perp)\Ph_0} &=& 0, \ {\rm for}\
\inprod{X^\perp}{X}=0, \\
\exp(2\pi d(X)) \Ph_0 &=& \Ph_0,
\eea
\ese
where the derivative of $V[\Ph]$ is defined in Eq.~(\ref{scderiv}).

Consider a vortex solution, defined by $(\Ph_0, X)$, satisfying
the above three conditions. Two specific gauge
choices have been made: firstly, it is in the temporal gauge; and,
secondly, the gauge is such that the solution is axially
symmetric. However, there is still residual gauge freedom defined from
the {\em global} gauge transformations of $G$.  

To track the gauge equivalence of solutions we shall compare solutions
that also share the same $\Ph_0$. Then the relevant group is the
subgroup of global gauge transformations defined by $G$ that leave
$\Phi_0$ invariant. Clearly this is defined by $H$.

Thus the gauge freedom of solution (\ref{eq-11a}, \ref{eq-11b}) is
given by the group $H$, with transformation:
\bse
\bea
\label{eq-16a}
\Ph(r,\theta) &\mapsto& D(h) \Ph(r,\theta) = f_{\rm NO}(X; r) 
D(\exp({\Ad}(h)X \theta))\Ph_0, \\
\label{eq-16b}
\underline{A}(r,\theta) &\mapsto& \ \ \ {\Ad}(h)\underline{A} \ \ = 
\frac{g_{\rm NO}(X; r)}{r} {\Ad}(h) X
\underline{\hat{\theta}}.
\eea
\ese
This is equivalent to transforming the pair: 
\beq
(X, \Ph_0) \mapsto ({\Ad}(h)X, \Ph_0).
\eeq

Hence, all the pairs $\{(\Ad(h)X, \Ph_0) : h \in H\}$ define vortex
solutions that are gauge equivalent. If two vortices may not be so
related then they are gauge inequivalent. 

The origin of family structure is now transparent. In general we write
\beq
\calG = \calH \oplus \calM,
\eeq
with the orthogonality defined by (\ref{scaleip}). This is the {\em
reductive decomposition} and has the following properties 
\beq
\label{red}
[\calH, \calH] \subseteq \calH\ \ \ {\rm and}\ \ \ 
 [\calH, \calM] \subseteq \calM.
\eeq
One interprets $\calM$ as the space of generators defining massive
gauge bosons. If $\calM$ is reducible under ${\Ad}(H)$, it may be
decomposed into orthogonal irreducible subspaces 
\beq
\label{reduct}
\calM = \calM_1 \oplus \ldots \oplus \calM_N.
\label{eq-19}
\eeq
This is the origin of family structure: vortices with generators lying
in the same $\calM_i$ may be gauge equivalent; whilst two vortices
with generators lying within different $\calM_i$'s may not be.

However, more importantly, the decomposition (\ref{reduct}) allows one
to classify all embedded vortex solutions:\\ 
(i) embedded vortices are defined by $X \in \calM_i$
\fn{the situation is slightly more complicated for those $\m_i$'s of
non-trivial rank, as we discussed in the prenote}.\\
(ii) there may be additional {\em combination} vortex solutions lying
within $\calM_i \oplus \calM_j$ for certain critical values of the
coupling constants such that 
\beq
\frac{\norm{d(X_i)\Ph_0}}{\norm{X_i}} = 
\frac{\norm{d(X_j)\Ph_0}}{\norm{X_j}},
\eeq
with $X_i \in \calM_i$, $X_j \in \calM_j$.

We shall discuss the implications of these results before discussing
the method of proof.

\subsection{Implications}

Consider a generator $X_i \in \calM_i$ normalised such that
$D(\exp(2\pi X_i))\Ph_0 = \Ph_0$. Furthermore let this generator be
minimal, so that $D(\exp(\theta X_i)) \Ph_0 \neq \Ph_0$ for $\theta
\in (0, 2\pi)$. The elements of this set all generate vortices of unit
winding number, and define the manifold
\beq
\Ad(H)X_i \subset \calM_i.
\eeq
When ${\rm rank}(\calM_i)=1$, $\Ad(H)$ is transitive over $\calM_i$
and this manifold is a $k$-sphere with $k=\dim(\calM_i)-1$. 

Similarly the corresponding vortex of winding number $n$ is generated
by $n X_i$. Thus when combination vortices are not admitted a
solutions, the total space of vortex generators is: 
\beq
\{\sum_{n_1 \neq 0} n_1 {\Ad}(H) X_1 \} \oplus \ldots
\oplus  \{ \sum_{n_N \neq 0} n_N {\Ad}(H) X_N \}.
\label{eq-20}
\eeq
This represents a collection of concentric maximal spheres
within each vector space $\calM_i$, with the radii of the spheres
being proportional to the winding numbers of vortices defined from
generators within that sphere. Note that if any ${\rm rank}(\calM_i)
>1$ then the situation becomes more complex.

This picture is useful when considering gauge equivalence of
vortices: gauge transformations of the form~(\ref{eq-16a}, 
\ref{eq-16b}) move vortex generators $X$ around the particular maximal 
sphere in which $X$ lies, taking $X \mapsto {\Ad}(h)X$. Such
transformations have two distinct types:\\ 
(i) when $\dim{\calM_i} = 1$ the space of vortex generators
corresponds to a set of equally spaced points. The corresponding
vortices are invariant under $H$. \\ 
(ii) when $\dim{\calM_i} > 1$ the space of vortex generators
corresponds to a set of concentric maximal spheres. The corresponding
vortices are not invariant under $H$, with two vortices within the
same sphere having generators related by $\Ad(H)$. Note that such
vortices are gauge equivalent to their anti-vortex. 

\subsection{Method of Proof}

For pedagogical reasons we shall delay a proper treatment of the
formalism and proof to another paper~\cite{me4}, and present here
instead a sketch of how the proof proceeds. 

We shall repeat again the conditions upon a generator $X$ of an
embedded vortex of the form~(\ref{eq-11a}, \ref{eq-11b}):\\
(i) The boundary conditions of the vortex form a closed geodesic on
the vacuum manifold, so that with appropriate normalisation
$\exp(2\pi X) \in H$.\\ 
(ii) In order that the embedded vortex is a solution to the equations
of motion one needs $\inprod{d(X)\Ph_0}{d(X^\perp)\Ph_0}= 0$ for all
$X^\perp$ such that $\inprod{X^\perp}{X}=0$.  

Only vortex generators that satisfy these two conditions define
embedded vortices. We shall firstly indicate how vortex generators in
$\calM_i$ are classified, before discussing combination vortices.

Vortex generators that satisfy (i) above are classified by the
following conjecture:
\begin{conjecture}
For $X \in \m_i$, the geodesic $\ga_X(\theta)=\{D(e^{X\theta})\Phi_0
:\theta \in {\bf R} \}$ defined from $X$ is closed. Note that we are
restricting $\calM_i$ such that ${\rm rank}(\calM_i)=1$.
\end{conjecture}
Proof of this result is discussed in~\cite{me4} for a wide variety of
cases by relating it to the embeddings of maximal Abelian subalgebras 
within $\calG$. These maximal Abelian subalgebras generate toroidal
submanifolds of the vacuum manifold, and the closed geodesics within
such toroidal submanifolds are easily found.

Vortex generators satisfying condition (ii) are categorised by the
following theorem:
\begin{theorem}
The (real) inner product on $\calV$ is related to the inner product on 
$\calM$ by 
\bea
\inprod{d(X_i)\Phi_0}{d(Y_j)\Phi_0} = \la_i \la_j
\inprod{X_i}{Y_j}, \ X_i \in \m_i, Y_j \in \m_j, \nn \\
{\rm where}\  \la_i =
\frac{\norm{d(X_i)\Phi_0}}{\norm{X_i}}. \nn  
\eea
Each $\la_i$ is constant upon its particular $\m_i$.
\end{theorem}
This theorem is proved in \cite{me4} by relating it to the conformal 
properties of inner products.

We now consider combination vortices. These are vortices with
generators lying between, say, $\calM_i$ and $\calM_j$. Generally the
generator of a combination vortex may be written $X = \al X_i + \be
X_j$, where $X_i \in \calM_i$ and $X_j \in \calM_j$. The spectrum of
these that close is generally rather complicated, and may be either a
continuous spectrum between $\calM_i$ and $\calM_j$ or a discrete one.

For $\inprod{d(X)\Ph_0}{d(X^\perp)\Ph_0} =0$ to be satisfied for
combination vortices one need only consider $X^\perp = X_i/(\al
\norm{X_i}^2) - X_j/(\be \norm{X_j}^2)$; then direct substitution
yields the following condition: 
\beq 
\frac{\norm{d(X_i)\Ph_0}}{\norm{X_i}} = 
\frac{\norm{d(X_j)\Ph_0}}{\norm{X_j}}.
\label{eq-21}
\eeq
This simple condition deserves some comment. The expression
$\norm{d(X)\Ph_0} / \norm{X}$ is independent of the generator $X
\in \calM_i$ (by scaling and adjoint action), and is thus dependant
only upon the coupling constants acting as scales in the inner 
product~(\ref{scaleip}). One thus interprets condition (\ref{eq-21})
as a restriction upon the ratios of gauge coupling constants that are
allowed in order to admit embedded combination vortex solutions.

%%%%%%%%%%%%       section 4      %%%%%%%%%%%%%%%%%%%%%%%%%%%%%%%%%

\section{Stability of Embedded Vortices}
\label{sec-4}

After determining the family structure, one wishes to determine how
this spectrum of vortices relates to the stability of embedded
defects. There are two types of stability to consider:\\ 
(i) Topological stability, quantified by the first homotopy group of
the vacuum manifold. Such vortices may either be Abelian or
non-Abelian as specified below.\\
(ii) Dynamical stability \cite{Vach91}, which occurs when the theory
admits stable semi-local vortices in a limit (the semi-local limit) of
the coupling constants. By continuity dynamical stability occurs for a
finite region of coupling constant space around the region of stable 
semi-local vortices. 

Firstly, we shall remind the reader of some concepts that will prove
useful to the following discussion. Recall that the centre $\calC$ of
$\calG$ is the set of elements that commute with $\calG$. Then the
stability of vortices is related to the projection of $\calC$ onto
$\m$,  
\beq
{\rm pr}_\calM(X) = X + X_h,
\eeq
with $X_h \in \calH$ the unique element such that ${\rm pr}_\calM(X)
\in \calM$. We shall prove later that ${\rm pr}_\calM(\calC)$ consists
of one-dimensional irreducible $\m_i$'s. 

The main result of this section is that: ${\rm pr}_\calM(\calC)$
differentiates between Abelian topologically stable or dynamically
stable embedded vortices. The distinction being:
\begin{enumerate}
\item Topologically stable Abelian vortices are generated by elements
in the {\em intersection} of $\m$ and $\calC$.
\item Dynamically stable vortices (in some coupling constant limit)
are generated elements in the non-trivial {\em projection} of $\calC$
onto $\m$. 
\end{enumerate}
Such vortices always correspond to one-dimensional $\calM_i$'s.

\subsection{Topologically Stable Abelian Vortices.}

Abelian topological vortices are present for a symmetry breaking 
of the form
\beq
G = G' \times U(1)_1 \x ... \x U(1)_N \rightarrow H \subseteq G'.
\eeq
So that the symmetry breaking is the product of $U(1)_1 \x ... \x
U(1)_N  \rightarrow {\bf 1}$ and $G' \rightarrow H$, with there being
no Abelian part of $G'$ with the subalgebra $u(1) \subseteq \m$. The 
first half of the symmetry breaking gives a non-trivial topology, with
the topological charges being elements of ${\bf Z}^N$.

The Lie algebra of the group $G$ decomposes under the adjoint action
of $H$ as
\beq
\calG = \calH \oplus \calM' \oplus \calM_1 \oplus ... \oplus
\m_N,
\eeq
where
\beq
\calM_i = u(1)_i.
\eeq
The individual spaces $\m_1 ,..., \m_N$ are clearly irreducible
under $\Ad(H)$ and are thus one-dimensional irreducible parts of
$\calM$. In addition, $\calM'$ may decompose further under $\Ad(H)$;
however, this is irrelevant for the present discussion. 

Therefore the generators of stable Abelian vortices $\m_1 \oplus
... \oplus \m_N$ lie within the {\em intersection} of the Lie algebra 
of the centre of $G$ with $\m$, {\em i.e.}
\beq
\m_1 \oplus ... \oplus \m_N = \m \cap \calC.
\eeq

\subsection{Topological Stability from $S/{\bf Z}_n$}

Topological stability can arise from elements outside the
centre of the group if discrete symmetries are involved. For example
when $G \rightarrow H$ such that  
\beq
\frac{G}{H} \cong \frac{S}{{\bf Z}_n} \x ...
\eeq
with $S$ a simply connected manifold of dimension larger than
one, then one has non-Abelian topologically stable vortex solutions.
The topological charges are given by the equivalence classes of
the first homotopy group
\beq
\pi_1 \left( \frac{\calM}{{\bf Z}_n} \right) = {\bf Z}_n.
\eeq
However, for $S$ not simply connect one may have a more complex
structure, see for instance $^3$He in \cite{me3}. 

\subsection{Dynamical Stability}

A vortex may be dynamically stable~\cite{Vach91} if there exists a
limit of the coupling constants $\{ q_i \}$ such that the model then
admits stable semi-local vortices ---  {\em i.e.} in that limit it has
a breaking of the form below in Eq.~(\ref{eq-22}).  We refer to such a
limit as the {\em semi-local limit}; in that limit 
some of the coupling constants vanish, so that the corresponding
symmetry becomes global. By continuity, vortices are dynamically
stable for some finite region in coupling constant space around the
semi-local limit. 

Preskill~\cite{Pres92} has shown how to construct semi-local
defects generally. For vortices, one needs a symmetry breaking scheme
of the form
\beq
G_{\rm global} \x U(1)_{\rm local} \rightarrow H,\ \ \ {\rm with}\ H
\cap U(1)_{\rm local} = 1,
\label{eq-22}
\eeq
where the suffices `global' and `local' represent non-gauge and gauged
symmetries, respectively. Requiring that this does not admit
topological vortices leads to the condition that $H \not\subset
G_{global}$. Note that Eq.~(\ref{eq-22}) needs only be a sub-part of a
more general symmetry breaking.

The main result of this section is motivated through observing that we
are trying to isolate part of the symmetry breaking and then take a
limit in which this part of the symmetry breaking resembles
Eq.~({\ref{eq-22}). That $H \not\subset G_{\rm global}$ implies that
one of the generators of $\calH$ is a linear combination of generators in
$\calC$ and $\calG_{\rm global}$. The other linear combination, which is
perpendicular to this, lies in $\calM$ and is the clear
contender for vortex generators generating dynamically stable
embedded vortices. 

The result is then: 
\begin{theorem}
Vortex generators $X \in {\rm pr}_\calM(\calC)$ such that $X \not\in
\calC$ define embedded vortices that 
are stable in a well defined semi-local limit of the model and are
thus dynamically stable for a region of parameter space around that
semi-local limit. 
\end{theorem}

\proof{
Consider $X \in {\rm pr}_\calM(\calC)$, such that it generates a
closed geodesic. Then $X = X_c + X_h$, with $X_c \in \calC$, such that
${\rm pr}_\calM (X_c) = X$, generating a $U(1)_c \subseteq C$ and
$X_h \in \calH$ generating a $U(1)_h \subset H$. These define a
decomposition of the symmetry breaking of the Lie algebras: 
\beq
\calG = \calG' \oplus u(1)_c \rightarrow 
\calH= \calH' \oplus u(1)_h,
\eeq
with $U(1)_c \cap H = 1$. It is now clear that the appropriate
semi-local limit to obtain a symmetry breaking of the form
Eq.~(\ref{eq-22}) is to make coupling constants appertaining to $G'$
vanish. The corresponding vortex generator is $X$. This completes the
proof.} 

We also have a useful subresult about the dimension of $\calM_i$'s for
such vortices:
\result{Lemma}{
Vortices with a stable semi-local limit are {\em always} generated by
generators in one dimensional $\calM_i$'s.
}

\proof{
The proof that if $\calM_i \subseteq {\rm pr}_\calM(\calC)$ then
$\dim(\calM_i)=1$ relies on the following identification: if
one writes $\calM_{\rm ker}$ as the collection of one dimensional
$\calM_i$'s then necessarily $\calM_{\rm ker} = \{X \in
\calM : {\rm \Ad}(H)X = X\}$. Now we are reduced to showing
${\rm pr}_\calM(\calC) \subseteq \calM_{\rm ker}$.

Consider $(X_c + X_h) \in {\rm pr}_\calM(\calC)$, with $X_c \in \calC$
and $X_h \in \calH$. By Eq.~(\ref{red}), $[\calH, X_c + X_h] \in
\calM$. But also $[\calH, X_c + X_h]=[\calH, X_h] \in \calH$.
Hence $[\calH, X_c +X_h]=0$, and equivalently $\Ad(H)(X_c + X_h)=X_c +
X_h$, proving the result.}

%%%%%%%%%%%%       section 5      %%%%%%%%%%%%%%%%%%%%%%%%%%%%%%%%%

\section{Extensions to Domain Walls and Monopoles}
\label{sec-5}

So far we have considered only how the properties of embedded vortices
are related to the underlying group theory of the formalism introduced
in~\cite{Vach94}. However, that formalism is also relevant to embedded
domain walls and embedded monopoles. Hence, we discuss in this section
how our approach may be extended to cover them. One should note that a
considerably expanded version of this discussion will be covered
in~\cite{me5}.

Non-topological embedded domain walls and embedded monopoles are
unstable. Non-topological embedded monopoles are unstable due to a
long range instability of the magnetic field \cite{Bran79, Cole};
which is related to the non-existence of semilocal
monopoles~\cite{Pres92}. Non-topological embedded 
domain walls are unstable due to a short range instability of the
scalar field~\cite{Vach94}. This contrasts strongly with the case of
non-topological embedded vortices, where one may still have dynamical
stability~\cite{Vach91, Pres92}

Thus we shall only discuss the nature of the family stability for
embedded domain walls and monopoles.

We deal with embedded domain walls and embedded monopoles separately. 

\subsection{Embedded Domain Walls}

An embedded domain wall solution is defined from the singleton
$\Ph_0\in \calV$: 
\bse
\bea
\Ph(x) &=& {\rm tanh}(cx) \Ph_0,\\
A^\mu &=& 0, 
\eea
\ese
where $c$ is some constant depending upon parameters in the
Lagrangian. This is related to the embedded subtheory formalism by
\bse
\bea
G_\emb &=& {\bf Z}_2, \\
H_\emb &=& {\bf 1}, \\
\calV_\emb &=& {\bf R}\Ph_0.
\eea
\ese
Hence, provided that $G$ is transitive over the vacuum manifold all
embedded domain walls are gauge equivalent. 

\subsection{Embedded Monopoles}

An embedded monopole solution is defined from the embedding
\bea
G &\rightarrow& H \nn \\
\cup &\ & \cup \\
SU(2) &\rightarrow& U(1). \nn
\eea
With the embedded solution specified by the triplet $(\Ph_0, X_1, X_2)
\in \calV \x \calM \x \calM$, with the following constraints on the
pair $(X_1, X_2) \in \calM \x \calM$: \\
(i) The pair $(X_1, X_2)$ consists of two members of an orthogonal
basis of an $su(2) \subset \calG$, thus
\beq
\norm{X_1} = \norm{X_2}, \ \ \ \ \inprod{X_1}{X_2}=0,
\eeq
and 
\beq
[X_1, [X_1, X_2]] \propto X_2, \ \ \ \ [X_2, [X_1, X_2]] \propto X_1.
\eeq
(ii) the embedded $SU(2)$ is such that $SU(2) \cap H = U(1)$, thus
\beq
[X_1, X_2] \in \calH.
\eeq
(iii) the generators are properly normalised so that, for $i=\{1,2\}$,
\beq
\exp(2 \pi X_i) \Phi_0 = \Phi_0.
\eeq

Then the corresponding monopole solution (for winding number $n=1$) is
a usual $SU(2)$ hedgehog configuration embedded in $G \rightarrow H$  
\bse
\bea
\underline{\Ph}(\underline{r}) &=& f_{\rm mon}(r) \underline{\hat{r}},\\
A^\mu_a(\underline{r}) &=& \frac{g_{\rm mon}(r)}{r}
\ep_{\mu a b} X_b, 
\eea
\ese
where $f_{\rm mon}(r)$ and $g_{\rm mon}(r)$ are the usual monopole
profile functions. Notationally, we are treating $\Ph$ to be a
vector within its corresponding embedded subtheory and we are using
$X_3 = [X_1, X_2]$. 

The corresponding embedded sub-theory is defined by the triplet
$(\Phi_0, X_1, X_2)$:
\bse
\bea
G_\emb &=& \exp({\bf R}X_1 \oplus{\bf R}X_2 \oplus{\bf R}[X_1, X_2]),
\\ 
H_\emb &=& \exp({\bf R}[X_1, X_2]), \\
\calV_\emb &=& {\bf R} D(G_\emb) \Ph_0. 
\eea
\ese

We now discuss the family structure of embedded monopoles by
relating them to the family structure of embedded vortices with the
following observation:
\bea
G &\rightarrow& H \nn \\
\cup &\ & \cup \nn \\
SU(2) &\rightarrow& U(1) \\
\cup &\ & \cup \nn \\
U(1) &\rightarrow& {\bf 1}. \nn
\eea
Thus  monopole solutions always contain embedded vortex solutions. 
Using the results on embedded vortices we may immediately infer
\beq
(X_1, X_2) \in \calM_i \x \calM_i.
\eeq
Therefore monopoles defined from different $\calM_i$'s are gauge
inequivalent. There could also be further family 
structure arising from gauge inequivalent monopole configurations with
{\em different} pairs $(X_1, X_2)$ within the same irreducible
subspace $\calM_i$. 

The structure of embedded monopole solutions is discussed more
fully in a forthcoming manuscript~\cite{me5}.

\bigskip
\bigskip

%%%%%%%%%%%%%%%%%%%%%%%%%%%%% conclusions  %%%%%%%%%%%%%%%%%%%

{\noindent{\Large{\bf Conclusions}}}

\nopagebreak

\bigskip

\nopagebreak

We conclude by summarising our main results:
\begin{enumerate}
\item In section (2) we expanded on the embedded defect formalism of
\cite{Vach94} by making the underlying group theory explicit.
\item In section (3) we showed that the family
structure of embedded vortices is dependent only upon the group theory
of symmetry breaking. We derived how so, in a prescriptive way that 
allows a determination of the family structure. 
\item In section (4) we related this family structure to stability.
Classifying stability into (Abelian and non-Abelian) topological
stability and dynamical stability, we showed how Abelian topological
stability comes from the {\em intersection} of the centre of the
gauge group with vortex generators. Correspondingly dynamical
stability originates from the non-trivial {\em projection} of the
centre. 
\item In section (5) we showed how the points (2) and (3) above can be
carried over to embedded domain walls and embedded monopoles. 
\end{enumerate}

\bigskip
\bigskip

%%%%%%%%%%%%%%%%%%%%%%%%%%%%% acknowledgements %%%%%%%%%%%%%%%%%%%

{\noindent{\Large{\bf Acknowledgements}}}

\nopagebreak

\bigskip

\nopagebreak

This work is supported partly by PPARC and partly by the European
Commission under the Human Capital and Mobility programme. One of us
(NFL) acknowledges EPSRC for a research studentship. 
We wish to thank N. Manton and 
T. Vachaspati for interesting discussions related to this work when it
was in a preliminary state of development; also T. Kibble and N. Turok 
for interesting discussions and important clarification when this
work was in an advanced state. We would also like to thank the
anonymous referee for many invaluable comments and corrections.

\bigskip
\bigskip

%%%%%%%%%%%%%%%%%%%%%%%%% appendix  %%%%%%%%%%%%%%%%%%%%%%%%%

{\noindent{\Large{\bf Appendix: On the Nielsen-Olesen Profile
      Functions}}} 

\nopagebreak

\bigskip

\nopagebreak

The Nielsen-Olesen Functions are the (scaled) profiles of the scalar
and gauge fields within the vortex solution. They depend upon the
scalar quartic self-coupling and the charge, through the parameters in
the Lagrangian describing the field theory. Although we have tried to
avoid referring to the actual underlying field theory in this paper,
it becomes necessary when discussing the profile functions.

The Nielsen-Olesen profile functions are defined from the vortex
solutions in the Abelian-Higgs model. The Abelian-Higgs model is
defined through the Lagrangian, which describes the interaction
between the Higgs field $\Phi \in {\bf C}$ and the gauge field $A_\mu
\in i{\bf R}^4$: 
\beq
\calL [\la, q; \Phi, A_\mu] =
\frac{-1}{4} F_{\mu \nu}^* F^{\mu \nu} + \frac{1}{2} D_\mu \Phi^* D^\mu
\Phi - \la ( \Phi^* \Phi - \et^2 )^2,
\eeq
where
\bse
\bea
D_\mu &=& \partial_\mu + q A_\mu, \\
F_{\mu\nu} &=& \partial_\nu A_\mu - \partial_\mu A_\nu. 
\eea
\ese
Substitution of the {\em ansatz} for the vortex (of winding number
$n$)
\bse
\bea
\Phi(r, \theta) &=& \eta f(\la, q, n; r) e^{in \theta},\\
{\underline A}(r, \theta) &=& g(\la, q, n; r) \left( \frac{in}{qr}
\right) \hat{\underline{\theta}}, \ \ \ \ A_0 = 0,
\eea
\ese
yields the Lagrangian for $f,g$:
\beq
\calL [\la, q; f, g] =
-\frac{n^2}{4q^2} \frac{1}{r^2} \left( \deriv{g}{r} \right)^2 -
\frac{\eta^2}{2} \left( \deriv{f}{r} \right)^2 -
\frac{n^2 \eta^2 f^2}{2r^2} (1 + g)^2 -\la \eta^4 (f^2 -1)^2.
\eeq
The Nielsen-Olesen profile functions $f_{\rm NO}$ and $g_{\rm NO}$
minimise this Lagrangian.

A useful identity
that relates the profile functions at different values of the
electric charge is:
\bse
\label{identity1}
\bea
f_{\rm NO}(\la, q', n; r) &=& f_{\rm NO}(\frac{\la}{(q'/q)^2}, q, n;
{\frac{q'}{q}}r),\\
g_{\rm NO}(\la, q', n; r) &=& g_{\rm NO}
(\frac{\la}{(q'/q)^2}, q, n; {\frac{q'}{q}}r).
\eea
\ese
Proof is provided by direct substitution into the Lagrangian.

In certain situations the profile functions for a general embedded
vortex may be related back to the Nielsen-Olesen profile functions
for the Abelian-Higgs model. Then one has recourse to the above
identity Eq.~(\ref{identity1}) to relate the profile functions between
different classes.

To examine the form of a general embedded vortex one needs to use the
Lagrangian for a general Yang-Mills theory. The Lagrangian is of the
form
\beq
\calL  =
\frac{-1}{4} \langle F_{\mu \nu}, F^{\mu \nu} \rangle + \frac{1}{2}
\langle D_\mu \Phi, D^\mu \Phi \rangle - V(\Phi),
\eeq
where
\bse
\bea
D_\mu &=& \partial_\mu + d(A_\mu), \\
F_{\mu\nu} &=& \partial_\nu A_\mu - \partial_\mu A_\nu - [A_\mu,
A_\nu].  
\eea
\ese
Here $\langle .,. \rangle$ are the respective inner products and $V(\Phi
)$ is the scalar potential, which is quartic in the components of
$\Phi$. Note that we are using the scaled representation, which
naturally introduces a coupling into the non-Abelian term of
$F^{\mu\nu}$ by constraining $d([A^\mu, A^\nu]) = [d(A^\mu), d(A^\nu)]$.

Consider the {\em ansatz} for an embedded defect solution
\bse
\bea
\Phi(r, \theta) &=& f(X; r) e^{d(X) \theta} \Phi_0,\\
{\underline A}(r, \theta) &=& \frac{g(X; r)}{r} X
\hat{\underline{\theta}}, \ \ \ \ A_0 = 0. 
\eea
\ese
Since $V(\Phi)$ is a scalar potential, and we implicitly assume that
the asymptotics of the embedded defect lie in the physical vacuum
manifold, which is the minimum of the potential, we must have:
\beq
V(\Phi(r, \theta)) = \la[\Phi_0] \eta^4 (f(r; X)^2 -1)^2,
\eeq
with $\eta = \norm{\Phi_0}$. We shall assume the vacuum manifold is
connected, then $\la$ is independent of $\Phi_0$ and $\eta$ is
independent of the direction of $\Phi_0$.

Elementary substitution of the {\em ansatz} into the Lagrangian yields
the Lagrangian describing the profile functions $f,g$:
\bea
\calL [\la, X ; f(X; r), g(X; r)] &=&
-\frac{\norm{X}^2}{4r^2} \left( \deriv{g}{r}\right)^2 -
\frac{\eta^2}{2} \left( \deriv{f}{r} \right)^2  \nn \\
-\norm{d(X) \Phi_0}^2 \frac{f^2}{2r^2} (1 + g)^2 &-&\la \eta^4
(f^2 -1)^2. 
\eea
and as before the Lagrangian is minimised by the Nielson-Olesen
profile functions $f_{\rm NO}$ and $g_{\rm NO}$.

Now, providing
\beq
\frac{\norm{d(X) \Phi_0}}{\norm{\Phi_0}} = n, \ {\rm with}\ n \in
  {\bf Z}, 
\label{condition1}
\eeq
the winding number of the defect, then we have a simple relation
between profile functions of embedded defects and the 
profile functions in the Abelian-Higgs model. One corresponds the
embedded defects profile functions to those of the Abelian-Higgs
model with charge and winding number:
\bse
\bea
q^2 &=& \frac{\norm{d(X)\Phi_0}^2}{\norm{\Phi_0}^2 \norm{X}^2}, \\
n &=&\frac{\norm{d(X)\Phi_0}}{\norm{\Phi_0}}.
\eea
\ese
We also have use of the identity (\ref{identity1}) for relating
embedded defects in different classes.
Providing the condition (\ref{condition1}) is satisfied, and the
winding number of two embedded vortices in different classes are equal
(the two vortices being defined by $X_1$ and $X_2$) then
\bse
\bea
f_{\rm NO}(\la,X_1; r) &=& f_{\rm NO}(\frac{\la}{\ga^2},X_2; \ga r),\\
g_{\rm NO}(\la,X_1; r) &=& g_{\rm NO}(\frac{\la}{\ga^2},X_2; \ga
r),\\ 
{\rm where}\ \ga &=& \frac{\norm{d(X_1) \Phi_0} \norm{X_2}}
{\norm{d(X_2) \Phi_0} \norm{X_1}}.
\eea
\ese

In the case of combination vortices (see Eq.~(\ref{eq-21})) the
factor $\ga$ is one (by definition) and the profile functions
naturally coincide. This is to be expected since in many cases the
vortices in different classes may be continuously deformed into one
another when combination vortices exist.

It would be useful to know when the above condition (\ref{condition1})
is generally satisfied. The condition does not seem to be true
generally, and one may construct realistic counter examples . However,
there are also many theories which act 
as examples, for example the Weinberg-Salam theory.

It should be noted that cond. (\ref{condition1}) is equivalent to the
following condition: 
\bea
e^{2 \pi d(X)} \Phi_0 = \Phi_0, \ \ \ {\rm and}\ X \in
\calM_i, \nn \\ 
{\rm and}\ \frac{\norm{d(X) \Phi_0}}{\norm{\Phi_0}} \in {\bf Z},
\eea
{\em i.e.} when do the generators act like real numbers upon
exponentiation? As a simple example the Pauli spin matrices do.

%%%%%%%%%%%%%%%%%%%%%%%%%%%  end of text   %%%%%%%%%%%%%%%%%%%%%%%

%bibliography

%end of document
\end{document}